# Vers une Approche Inclusive de la Responsabilité Sociale des Entreprises (RSE) au Maroc : L'Engagement de la CGEM

Towards an Inclusive Approach to Corporate Social Responsibility (CSR) in Morocco: CGEM's Commitment.

Auteur 1 : GNAOUI Imane
Auteur 2 : MOUTAHADDIB Aziz

**Imane GNAOUI,** Doctorante
Ecole Nationale de Commerce et de Gestion, Université Ibn Tofail
Laboratoire de Gestions des Organisations

**Aziz MOUTAHADDIB,** Professeur d'Enseignement Supérieur
Ecole Nationale de Commerce et de Gestion, Université Ibn Tofail
Laboratoire de Gestions des Organisations

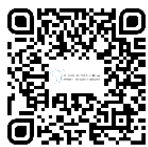
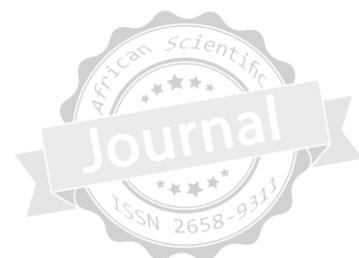

**Résumé :**

La responsabilité sociale des entreprises encourage les entreprises à intégrer des préoccupations sociales et environnementales dans leurs activités ainsi que dans leurs relations avec les parties prenantes. Elle englobe l'ensemble des activités visant le bien social au-delà des intérêts de l'entreprise et des exigences légales.

Divers organismes internationaux, auteurs et chercheurs ont exploré la notion de RSE et ont proposé une panoplie de définition reflétant leurs perspectives sur le concept.

Au Maroc, bien que les entreprises marocaines ne se basculent pas face aux portes de la RSE, plusieurs facteurs y incitent à intégrer la démarche RSE, non seulement dans leurs discours, mais également dans leurs stratégies.

La CGEM s'engage activement dans la promotion de la RSE au sein des entreprises marocaines en attribuant le « Label CGEM pour la RSE » aux entreprises respectant les critères mentionnés et énoncés dans la charte RSE. Le processus de labélisation des entreprises marocaines est en pleine expansion. Les graphes présentés dans cet article sont ventilés selon plusieurs critères tels que : la taille de l'entreprise, le secteur d'activité, la cotation à la Bourse de Casablanca, afin de fournir une vue d'ensemble sur les entreprises labellisées RSE au Maroc.

L'approche adoptée pour cet article est une approche qualitative visant à présenter, dans un premier temps, les différentes définitions du concept RSE et son évolution au fil du temps. Ainsi, l'étude se focalise sur le contexte marocain pour décortiquer et analyser l'état d'avancement de l'intégration de la RSE au Maroc et les différents efforts déployés par la CGEM pour sa mise en œuvre. Selon les données, 124 entreprises marocaines labellisées RSE. Pour un label existant depuis 2006, ce chiffre témoigne d'une certaine réticence des entreprises marocaines à mettre en œuvre pleinement la démarche RSE dans leurs stratégies. Néanmoins, le Maroc se trouve dans une phase de transition, marquée par l'adoption progressive des diverses pratiques socialement responsables.

**Mots Clés :** Responsabilité sociale des entreprises ; Confédération Générale des Entreprises Marocaines ; INDH ; Charte RSE ; Label CGEM.

**Abstract**

Corporate social responsibility encourages companies to integrate social and environmental concerns into their activities and their relations with stakeholders. It encompasses all actions aimed at the social good, above and beyond corporate interests and legal requirements.

Various international organizations, authors and researchers have explored the notion of CSR and proposed a range of definitions reflecting their perspectives on the concept.

In Morocco, although Moroccan companies are not overwhelmingly embracing CSR, several factors are encouraging them to integrate the CSR approach not only into their discourse, but also into their strategies.

The CGEM is actively involved in promoting CSR within Moroccan companies, awarding the "CGEM Label for CSR" to companies that meet the criteria set out in the CSR Charter. The process of labeling Moroccan companies is in full expansion. The graphs presented in this article are broken down according to several criteria, such as company size, sector of activity and listing on the Casablanca Stock Exchange, in order to provide an overview of CSR-labeled companies in Morocco.

The approach adopted for this article is a qualitative one aimed at presenting, firstly, the different definitions of the CSR concept and its evolution over time. In this way, the study focuses on the Moroccan context to dissect and analyze the state of progress of CSR integration in Morocco and the various efforts made by the CGEM to implement it. According to the data, 124 Moroccan companies have been awarded the CSR label. For a label in existence since 2006, this figure reflects a certain reluctance on the part of Moroccan companies to fully implement the CSR approach in their strategies. Nevertheless, Morocco is in a transitional phase, marked by the gradual adoption of various socially responsible practices.

**Introduction :**

L'ultime vocation des entreprises était de réaliser un profit pour satisfaire les besoins de ses actionnaires, selon Friedman en 1962 : « La seule responsabilité de l'entreprise est de faire du profit pour rémunérer ses actionnaires… »

En 1984, l'émergence de la théorie des parties prenantes de Freeman, marque un tournant majeur dans la perception de l'entreprise. Cette théorie met en avant non seulement l'intérêt accorder aux actionnaires mais aussi aux autres parties prenantes, passant ainsi d'une vision centrée sur les actionnaires « Shareholders » vers une celle qui intègre les intérêts de tous les « Stakeholders ».

Les entreprises se sont orientées progressivement vers l'intégration des pratiques socialement responsables, tout en reconnaissant l'importance de réalisation de bénéfice et de création du profit, sans nier la nécessité de répondre aux attentes des autres parties prenantes. La responsabilité sociale des entreprises représente un principe selon lequel les entreprises intègrent l'ensemble des préoccupations sociales, environnementales dans leurs activités commerciales. L'objectif est de transformer l'entreprise en un modèle de citoyenneté exemplaire répondant aux besoins de ses parties prenantes telles que les clients, actionnaires, fournisseurs, communauté locale et la société en général.

Les organismes internationaux, les auteurs et les chercheurs se sont tous penchés sur la notion de responsabilité sociale des entreprises afin de l'explorer. Plusieurs acceptions, interprétations, définitions et approches ont été évoqué, ce qui a créé un éventail de perspective sur le concept de RSE. Chaque définition reflète la vision et la perception spécifique des auteurs sur la notion et sur la manière d'intégration des entreprises des préoccupations sociales dans leurs activités commerciales.

Dans le contexte marocain, on ne peut pas nier que les entreprises marocaines ne se précipitent pas vers les portes de la RSE, malgré qu'elles se trouvent dans un contexte mondial globalisé et de plus en plus axé sur les enjeux de l'environnement et de developpement durable. Cependant plusieurs facteurs incitent les entreprises vers l'intégration de la RSE dans leurs stratégies à savoir : des dirigeants ambitionnés en quête de modernité et d'innovation prêts à explorer des nouvelles approches, l'implantation des filiales étrangères engagées dans des pratiques RSE alignées sur les exigences de leurs sociétés mères ainsi que la volonté royale incitant les entreprises d'adopter les bonnes pratiques commerciales.

Le pouvoir public marocain a ainsi procédé à l'institutionnalisation de la RSE et du développement durable par la signature de plusieurs conventions internationales, création d'une panoplie d'organisme spécialisé dans tous ce qui est d'ordre social, économique et environnemental et la mise en place de plusieurs initiatives à savoir l'INDH, Charte et Label RSE, création de la norme nationale de conformité sociale…etc.

Cet article adopte une approche qualitative basée sur une revue de littérature, à travers laquelle nous analyserons le cadre théorique et institutionnel de la notion de RSE, en explorant ses différentes définitions ainsi que son évolution au fil du temps. Nous recentrerons notre étude sur la mise en œuvre de la RSE au Maroc. En utilisant la technique du recueil documentaire, nous collecterons les données et les informations issues des écrits et documents déjà existant afin d'étudier l'état d'avancement de la démarche RSE au sein du pays, et obtenir une vision actuelle et actualisée de son intégration dans le tissu économique marocain à travers la présentation d'un certain nombre de graphique. L'objectif est de fournir une perspective globale de la Responsabilité Sociale des entreprises, dans un cadre institutionnel et théorique ainsi que le rôle de la Confédération Générale des Entreprises au Maroc dans la promotion de la RSE au sein de l'économie Marocaine.

Cette étude permet de répondre à la problématique suivante : Comment peut-on définir la Responsabilité Sociétale des Entreprises (RSE), quel est le niveau d'avancement de son intégration au sein des entreprises marocaines, et quel rôle joue la CGEM dans ce processus ?

Dans la première partie de l'article « I. RSE au Maroc », nous explorerons le cadre théorique et institutionnel de la RSE ainsi que la structure institutionnelle de la RSE dans le contexte marocain.

Dans la deuxième partie « II. Engagement de la CGEM pour la promotion de la RSE », nous commencerons par présenter la Confédération Générale des Entreprises au Maroc, ensuite nous examinerons ses initiatives visant à promouvoir la RSE telles que le Label « Label CGEM pour la RSE » et la charte RSE. Enfin, nous analyserons la dynamique de labellisation des entreprises au Maroc.

## I. RSE au Maroc :

### 1. Cadre théorique et institutionnel de la RSE :

Plusieurs acceptions ont été avancées, plusieurs orientations et approches sont prises par les chercheurs pour cerner le concept de Responsabilité sociale des entreprises, sans pour autant se dégage un consensus.

La Responsabilité sociale des entreprises remet en question l'idée que le profit est l'objectif unique de l'entreprise. Bien que ces dernières prennent en considération leurs environnements économiques au sens large, l'intégration de la RSE qui permet de s'intéresser à 3 grands types de facteurs : environnementaux, sociaux et de gouvernance, reste essentielle et demeure une nécessité.

Au début du $20^{ème}$ siècle, Henry Ford considéré l'entrepreneur pionnier de la RSE. Bien que le terme n'ait été largement utilisé et répandu à son époque, il a transmis ces valeurs en introduisant les concepts de paternalisme et de philanthropie.

Le paternalisme de Henry Ford se manifestait principalement par ses politiques en faveur de ses employés. En 1914, la plus célèbre action de paternalisme était « Ford $5 Day » qui consistait à augmenter le salaire minimum de ses ouvriers à 5 dollars par jour, dans le but d'améliorer les conditions de vie de ses employés et réduire le taux de rotation (turnover). Ainsi qu'il a investi et participé dans des initiatives philanthropiques qui ont eu un impact social plus large à savoir l'éducation et formation (fondation des écoles, des centres de formation…), santé et bien-être (services médicaux et sociaux, soins de santé…).

L'origine de la notion de RSE remonte à l'expression Anglo-américaine « Corporate social responsability », qui a vu le jour aux États-Unis en 1950 en raison des préoccupations religieuses et éthiques. L'objectifs est d'octroyer un espace aux actions et initiatives philanthropiques visant à agir sur le bien commun dans le but de promouvoir des causes sociales et humanitaires.

Les années 50-60 ont marqué le début de la notion de RSE. Howard R. Bowen est une figure clé dans l'histoire de la RSE à cette époque. Il a établi une définition en se basant sur le recensement des pratiques de la RSE dans son livre *« Social Responsability of the Businessman »* en 1953. Son ouvrage a joué un rôle crucial en montrant que les entreprises sont soumises à des responsabilités sociales et éthiques qui vont au-delà de leurs obligations économiques et légales.

<u>Les années 1970-1980</u> ont été marqué par l'émergence des préoccupations environnementales et sociales, en particulier avec le premier Jour de la Terre en 1970, qui a été un événement emblématique qui a mobilisé des millions de personnes aux États-Unis et à contribuer à sensibiliser le public aux problèmes environnementaux.

Vers la fin des années 1980, le concept de « Développement durable » a émergé comme une nouvelle approche pour répondre aux défis économiques, sociaux et environnementaux. Le rapport de Brundtland, publié en 1987 par la commission mondiale sur l'environnement et le developpement des Nations Unies a été l'élément clé et l'action majeure dans la popularisation de cette notion. Ce concept novateur a remplacé la notion précédente de « Eco developpement » par une approche holistique et intégrée, prenant en considération à la fois les aspects économiques, sociaux et environnementaux.

De surcroit, la théorie des parties prenantes adoptée par Edward Freeman dans les années 1980 a élargi la perspective et la vision de la RSE en mettant en lumière l'idée que les entreprises disposent des responsabilités envers toutes les parties prenantes (client, fournisseurs, employé…etc.) et non uniquement envers les actionnaires.

Les scandales financiers d'entreprise ainsi que les crises économiques et environnementales, ont mis en lumière les lacunes et les insuffisances des pratiques commerciales traditionnelles, incitant les entreprises à adopter des approches plus responsables vers les années 1990-2000. En conséquence, les initiatives de RSE se sont multipliées, incluant des pratiques commerciales durables, transparentes et éthiques dans leur stratégie et leurs opérations.

Le fameux scandale financier d'Enron, une société américaine de négoce d'énergie et de services publics fondée en 1985, était suite à l'utilisation des pratiques comptables créatives et des structures financières complexes afin de dissimuler ses pertes et embellir ses résultats financiers. En 2001, les irrégularités comptables d'Enron ont été révélées entrainant une chute spectaculaire de la valeur des actions de l'entreprise et perte de la confiance des investisseurs. En déposant son bilan, l'entreprise devient l'une des plus grandes faillites de l'histoire des Etats-Unis à l'époque. Le scandale d'Enron a mis en lumière les dangers de la fraude comptable, de la culture de l'avidité et de l'absence de surveillance adéquate dans les entreprises.

En réponse à ce scandale et pleins d'autres à l'époque, le gouvernement américain a adopté des réformes législatives importantes pour renforcer la réglementation financière et la surveillance des entreprises tel que la Loi Sarbanes-Oxley, adoptée en 2002, imposant des strictes matières de transparence financière, de gouvernance d'entreprise et de responsabilité des dirigeants.

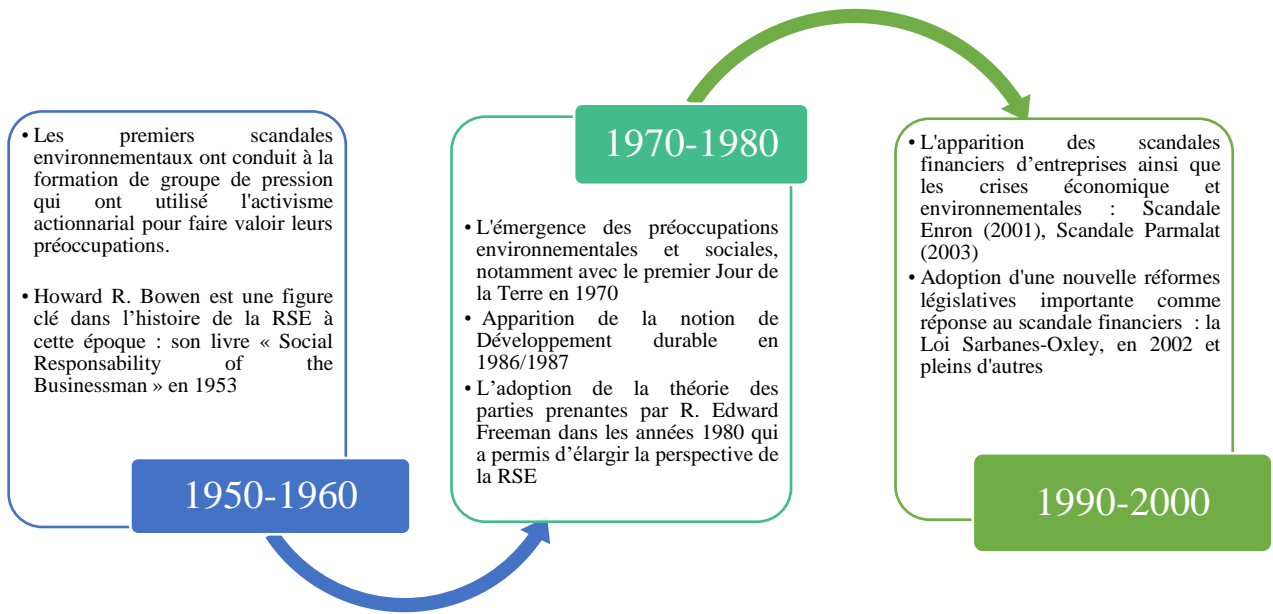

**Figure 1 : Evolution de la construction théorique de la RSE (1950-2000)**

*Source : Réalisé par nos propres soins*

❖ **Définition institutionnelle de la RSE :**

Les organismes internationaux attachent une grande importance à la responsabilité sociale des entreprises (RSE). Chacune d'eux propose une définition distincte en fonction de sa propre perspective. Le tableau ci-dessous, présente les différentes définitions fournies par chaque organisme international.

**Figure 2 : Tableau regroupant les définitions institutionnelles de la RSE**

| | Définition de la RSE |
|---|---|
| | **Définition institutionnelle de la RSE** |
| **Organisme** | **Définition** |
| **Commission Européenne (Livre Vert (2001))** | Le concept de RSE est défini comme : "L'intégration volontaire par les entreprises des préoccupations sociales et environnementales à leurs activités commerciales et leurs relations avec leurs parties prenantes"<br>Il s'agit d'aller au-delà des obligations légales pour investir dans le capital humain, l'environnement et les relations avec les parties prenantes.<br><br>Deux dimensions de la RSE :<br><br>♣Dimension interne : Concerne les pratiques internes des entreprises, telles que la gestion des ressources humaines, la santé et la sécurité au travail<br><br>♣ Dimension externe : Concerne l'interaction avec les parties prenantes externes telles que les communautés locales, les partenaires commerciaux, les fournisseurs... |
| **Union Européenne** | "Le concept de RSE signifie essentiellement que celles-ci décident de leur porpre initiative de contribuer à améliorer la société et rendre plus propres l'environnement. Cette responsabilité s'exprime vis-à-vis des salariés et plus généralement, de toutes les parties prenantes qui peuvent, à leur tour, influer sur sa réussite" |
| **Business for social responsability** | La CSR est définie comme le fait que l'entreprise se comporte de manière à satisfaire ou dépasser les attentes éthiques, légales, commerciales et publiques que la société manifeste envers les entreprises. |
| **World Business Council on Sustainable Development** | La CSR est l'engagement de l'e/se à contribuer à un développement durable, en travaillant avec ses employés, leur famille, la communauté locale et la société dans son ensemble pour améliorer la qualité de vie<br><br>La RSE consiste en un engagement des e/ses à agir dans un cadre légal en vue de participer au progrès économique et de contribuer à l'amélioration de la qualité et de la société dans son ensemble. |
| **AFNOR [Association Française de Normalisation] (2006)** | La RSE est : "la responsabilité d'une organisation vis-à-vis des impacts de ses décisions et activités sur la société et sur l'environnement, se traduisant par un comportement éthique et transparent qui :<br><br>♦ Contribue au développement durable, à la santé et au bien-être de la société<br>♦Prend en compte les attentes des parties prenante;<br>♦Respecte les lois en vigueurs et qui est en accord avec les normes internationales de comportement;<br>♦est intégré dans l'ensemble de l'organisation et mis en œuvre dans ses relations |

*Source : Réalisé par nos propres soins-Inspiré des articles et ouvrages*

❖ **Définition théorique de la RSE :**

Les auteurs accordent une attention particulière à ce concept de la RSE, ce qui explique la diversité et variété des définitions et des conceptualisations de cette notion. Au fil de temps, la RSE a connu une évolution passant des définitions restreintes à des acceptions plus larges.

Le tableau ci-dessous regroupe l'ensemble des définitions attribuées à la RSE fournies par divers chercheurs et auteurs au fil des années.

**Figure 3 : Tableau regroupant les définitions théoriques de la RSE**

| Définition de la RSE | |
|---|---|
| Définition théorique de la RSE | |
| Auteurs | Définition |
| BOWEN (1953) | "La CSR renvoie à l'obligation pour les hommes d'affaires d'effectuer des politiques, de prendre les décisions et de suivre les lignes de conduite répondant aux objectifs et aux valeurs qui sont considérées comme désirable dans notre société" <br><br> Au plan académique, Bowen (1953) a ouvert le débat sur la RSE en évitant d'enfermer le concept dans une définition trop étroite. Il aborde la RSE comme obligation pour les chefs d'entreprise de mettre en oeuvres des stratégies, de prendre des décisions et de garantir des pratiques qui soient compatible avec les objectifs et lesvaleurs de la communautés de façon général. |
| Frederick W (1960) | "La responsabilité sociétale est la Volonté de voir que les ressources (humaine et économiques) sont utilisées à de larges fin sociales et pas simplement pour l'intérêt limité de personnes privées et de firmes" |
| Davis (1973) | "La CSR renvoie à la prise en considération par l'entreprise de problèmes qui vont au dela de ses obligations économique, technique et légales étroites ainsi qu'aux réponses que l'entreprise donne à ces problèmes. Cela signifie que la SR débute là où s'arrête la loi. " |
| Carroll A (1979) | Il intervient pour compléter la définition de Bowen et propose un modèle conceptuel reposant sur 3 dimensions essentielles à la RSE : Les principes de responsabilité sociale, la manière dont l'entreprise mets ses principes en pratique (sensibilité sociale) et les valeurs sociétales qu'elle porte. <br><br> " La RSE englobe les attentes économiques, légales, éthiques et discrétionnaires que la société à des organisations à un moment donné" <br> Pour Carroll les responsabilités sont présentées sous forme d'une pyramide à quatre niveaux montrant que les responsabilités économiques représentent la base et la condition de toute démarche de RSE. Ainsi au fur et à mesure que l'on monte dans la pyramide l'étendue des responsabilités diminue et le niveau le plus haut représente le domaine où la responsabilité sociale s'exprime d'une manière claire. Cela montre en conséquence que celle-ci relève davantage de la volonté et de la discrétion des e/ses. |
| Jones T (1980) | "La RSE est la notion selon laquelle les entreprises ont une obligation envers les acteurs sociétaux autres que les actionnaires et au-delà des prescriptions légales ou contractuelles" |
| Wartick et Cochran (1985) | Ils élargissent cette approche en mettant en exergue la spécificité de la RSE comme résultante de l'intéraction de 3 dimensions : Principes/processus/politiques <br><br> "Les responsabilités sociales sont déterminées par la société, et les tâches de la firmes sont : <br> ♣ Identifier et analyser les attentes changeantes de la société en relation avec les responsabilités de la firmes <br> ♣ Déterminer une approche globale pour être responsable face aux demandes changeante de la société <br> ♣ Mettre en œuvre des réponses appropriées aux problèmes sociaux pertinents. |
| Wood D (1991) | "La signification de la responsabilité sociale ne peut être appréhender qu'à travers l'interaction de 3 principes : la légitimité, la responsabilité publique et le discrétion managériale <br> Ces principes résultant de la distinction de 3 niveaux d'analyse : institutionnel, organisationnel et individuel. " <br><br> •**Le niveau institutionnel :** repose sur les principes de légitimité : les entreprises doivent répondre aux attentes de la société en respectant les normes et les valeurs sociales en vigueur. <br> •**Le niveau organisationnel :** repose sur le principe de la responsabilité publique : Les entreprises doivent intégrer les valeurs sociales dans leur mission, leur vision et leur culture d'entreprise. <br> •**Le niveau individuel** : qui repose sur le principe de la volonté managériale des dirigeants ; |
| McWilliams et Siegel (2000) | La RSE peut être considérée aussi comme étant "L'ensemble des actions visant le bien social au dela des intérêts de la firme et de ce qui est demandé par la loi" |
| Igalens (2003) | Il a évoqué l'avènement d'un nouveau paradigme avec la responsabilité sociale de l'entreprise. <br> La RSE peut alors être envisagée comme le concept managérial du développement durable. |
| Allouche et alii (2004) | Adopter un comportement de responsabilité sociale : "C'est répondre à la nécessité de maximiser les objectifs de l'entreprises par l'entremise de sa rentabilité, au profit toujours de l'actionnaire mais aussi de ses autres partenaires. |
| Allouche et Laroche (2004) | La responsabilité sociétale des entreprises s'impose comme un concept dans lequel les entreprises intègrent sur une base volontaire les préoccupations sociales, environnementales et économiques dans leurs activités et dans leurs interactions avec les différentes parties prenantes. <br> La Responsabilité Sociétale de l'Entreprise apparaît ainsi comme la déclinaison pour l'entreprise des concepts de développement durable qui intègrent les trois piliers environnementaux, sociaux et économiques. |
| Cazel (2006) | La RSE est : "protéiforme, polysémique, floue, imprécise, confuse, autant d'adjectifs dont on affuble fréquemment la notion de responsabilité sociale (ou sociétale) d'entreprise (RSE). Pourtant si la notion est sans doute d'origine scientifique, elle a largement débordé les cercles académiques pour se répandre dans le grand public, devenant également une notion profane » |

*Source : Réalisé par nos propres soins-Inspiré des articles et ouvrages*

**Dans une approche axée sur les auteurs**, la RSE est identifiée par sa capacité à satisfaire les besoins de l'ensemble des parties prenantes. **Selon Freeman (1984)** la RSE consiste à satisfaire et répondre aux attentes du groupe ou des individus qui peuvent affecter ou être affecter par les actions de l'organisation.

**Une approche axée sur le contenu** souligne que la RSE favorise l'intégration de multiple dimensions sociétales et sociales. **Husted (2000)** suggère que la RSE fait face à une logique de contingence, puisqu'elle est une fonction d'interaction entre les problèmes sociaux d'une part et la stratégie et la structure organisationnelle d'autre part, lesquelles sont intrinsèquement liées aux problèmes.

**Capron et Quairel Lanoizelée (2004)** évoque une approche mettant l'accent sur la déclinaison affirmant que le RSE est une déclinaison du developpement durable dans les entreprises, du fait qu'elle est bornée au niveau des entités en tant qu'application des valeurs et principes du developpement durable.

Parmi les multiples définitions attribuées à la notion RSE, celle de **Caroll (1979) et Wood (1991)** ont été les plus exploitées.

Wood (1991) complète l'étude de Carroll, et présente la RSE par le biais de 3 Niveaux : Le niveau institutionnel, organisationnel et individuel.

Carroll (1979) a montré la RSE sous forme d'une pyramide, stipulant que la responsabilité économique représente la base et la condition de toute démarche. Au fur et à mesure que l'on progresse dans la pyramide l'étendue et l'intensité des responsabilités diminue pour aboutir à la responsabilité philanthropique.

**Figure 4 : La pyramide de RSE selon Carroll en 1979**

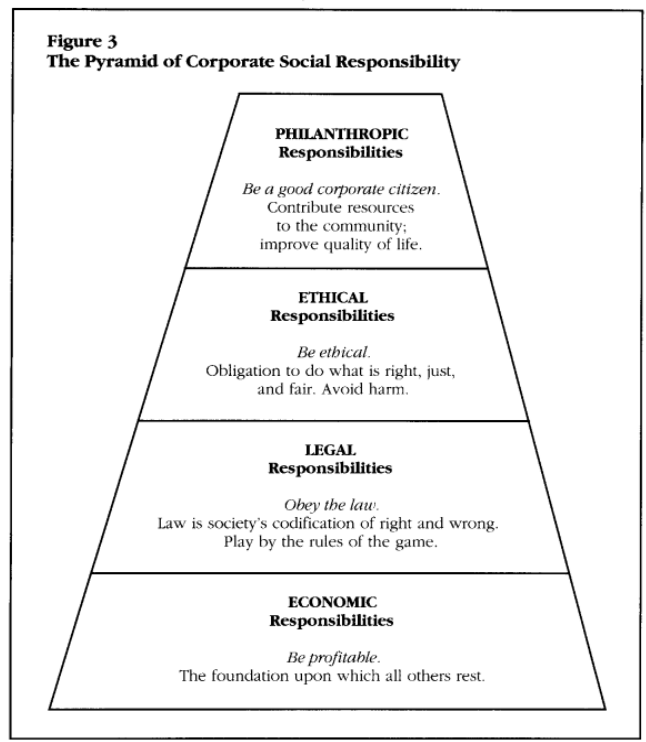

Source: *Archie B Carroll, "The Pyramid of Corporate Social Responsibility: Toward the Moral Management of Organizational Stakeholders", Article in Business Horizons · July 1991*

## 2. Cadre institutionnel de la RSE au Maroc :

Plusieurs facteurs expliquent l'engouement des entreprises marocaines et leurs orientations vers l'intégration de la démarche RSE aussi bien dans leurs discours que dans leurs stratégies.

En effet, la situation géographique du Maroc, par rapport aux principaux marchés internationaux, lui ont permis de s'ouvrir davantage sur l'économie mondiale d'où les entreprises marocaines se trouvent face à un ensemble de normes internationales et doivent s'aligner sur les bonnes pratiques.

Ainsi, les dirigeants des entreprises ont toujours ambitionné la modernité, et cherche à être toujours sur la pointe de l'innovation. Ceci les incite à non seulement se contenter des normes établies mais aussi de réinterpréter et réévaluer l'ensemble des comportements sociaux et de s'intéresser à des nouvelles approches, afin de les comprendre et les insérer de façon positive dans leurs stratégies.

En outre, les entreprises filiales des grands groupes internationaux implantées aux Maroc, se trouvent engager dans la démarche RSE en réponse à l'engagement de leurs société mère.

Enfin, la volonté royale qui encourage toute initiative socialement responsable reste une raison, entre autres, de l'enthousiasme des entreprises marocaines à la RSE. En 2005, lors de l'Intégrale d'investissement, qui a pour objectif de renforcer l'attractivité du pays, de favoriser les échanges et promotion des décisions de l'investissement, d'encourager les développements économiques pour les investisseurs étrangers et nationaux, le discours royal était un élément clés et un point fort et positif incitant les entreprises marocaines à mettre en place les outils de la responsabilité sociale ainsi que les indicateurs adéquats.

Le souverain affirmait avec solennité :

*« La responsabilité sociale ne saurait se réduire à de la compassion charitable, dès lors qu'elle est une condition essentielle de la viabilité, de la rentabilité à long terme et de l'acceptabilité sociale des investissements et de la croissance économique. C'est dans cet esprit que Nous suivons, avec le plus haut intérêt l'émergence d'un mouvement d'investissements et de placements financiers couplant les objectifs légitimes de rentabilité et de profits à des critères, non moins légitimes et universels, de responsabilité sociale et de développement humain et durable. Le Maroc, par sa législation et ses choix politiques et sociétaux, peut et veut être pour les investisseurs socialement responsables, un partenaire et une destination assumant pleinement les standards sociaux, environnementaux et de bonne gouvernance les plus avancés »*[1]

Les Principales mesures prises par les pouvoirs publics au Maroc en matière de RSE se manifestent par la signature de plusieurs conventions internationales et plusieurs protocoles. On peut citer, entre autres :

- **Convention sur le volet environnemental internationale** : la convention Marpol pour la prévention de la pollution par les navires en 1973, la convention des Nations Unies sur la lutte contre la désertification en 1994.

- **Conventions sur le Volet Social :** les conventions des Nations Unies dont le Pacte mondiale international relatif aux droits économiques, sociaux et culturels et le pacte mondiale des droits civils et politiques en 1979, la convention de l'organisation internationale du travail (OIT) en 1957.

---

[1] Discours Royale, en 2005, lors de l'intégral d'investissement (Site web : Message de SM le Roi à la troisième édition des "Intégrales de l'Investissement" | Maroc.ma)

- **Convention en relation avec la gouvernance :** Convention de l'Union Africaines sur la prévention et la lutte contre la corruption (2003), Convention Arabe contre la corruption (CACC) en 2010...

De surcroit, des mesures institutionnelles prises par le Maroc reflètent son intérêt à s'inscrire dans la philosophie du développement durable et à s'impliquer dans la démarche de RSE à savoir, la création de l'Initiative Nationale pour le Développement Humain (INDH) en 2005, l'investissement dans les secteurs des énergies renouvelables, la création d'une norme nationale de conformité sociale en 2010 intitulé l'Institut Marocaine de Normalisation (IMANOR) et lancement de la charte de la CGEM en 2006.

Ainsi que le Maroc dispose d'une panoplie d'organismes spécialisés dans tout ce qui est d'ordre social, économique et environnemental tels que : L'instance Centrale de Prévention de la Corruption (ICPC) en 2007, le conseil de la concurrence (CC) en 2009, le Conseil National des droits de l'Homme (CNDH) en 2011, l'observatoire de la Responsabilité sociale au Maroc (ORSEM) en 2017 et le Conseil économique social et environnemental (CESE) en 2021.

Les principales initiatives faites par le Maroc en faveur de la RSE et du Développement Durable se manifeste par la mise en place de l'Initiative Nationale pour le Développement Durable en 2005 par sa majesté le Roi Mohammed VI, qui est un moyen efficace d'encourager l'intégration des objectifs sociaux dans les choix des investissements, tout en luttant contre la pauvreté, la précarité et l'exclusion sociale. Cette initiative a bénéficié d'une enveloppe de 10 milliards de dirhams entre 2006 et 2010.

Le Maroc a assisté à une deuxième phase de l'INDH, qui s'étale de 2011 à 2015 avec de nouvelles perspectives prometteuses tant sur le plan de la forme que du fond. Cette seconde phase se caractérise par 5 principaux programmes [2]:

- Le 1$^{er}$ programme de lutte contre la pauvreté en milieu rural ; ayant pour objectif global d'améliorer la qualité de vie de la population rurale.

- Le 2$^{ème}$ programme de lutte contre l'exclusion sociale en milieu urbain dont les objectifs est de lutter contre l'exclusion sociale et d'améliorer les conditions et la qualité de vie de la population.

---

[2] INDHProgram2011-2015-FRENCH.pdf

- Le 3ᵉᵐᵉ programme de lutte contre précarité afin d'améliorer la qualité de vie des personnes précaires et soutenir les populations en situation difficile

- Le 4ᵉᵐᵉ programme transversale se base sur deux axes, l'axe Accompagnement et l'axe d'activité génératrices de Revenus.

- Le 5ᵉᵐᵉ programme visant la mise à niveau territorial via 5 axes d'intervention : appui à la santé (amélioration des offres de soins de base), électricité rurale, l'eau potable, routes et pistes rurales et enfin l'appui à l'éducation.

Cependant le Maroc, présente un IDH moyen selon le Rapport mondial sur le Développement Humain (2023/2024) publié par le « Programme des Nations Unies pour les Développement » (PNUD)³ . Les graphes ci-dessous, illustre l'évolution depuis 1990 jusqu' à 2022 de l'indice de développement Humain (IDH) **(Figure 5),** un indicateur utilisé pour évaluer les développements globaux des pays en se basant sur 3 dimensions principales : Santé (mesuré par l'espérance de vie), éducation (évalué par la durée moyennes de scolarisation des adultes et la durée attendue de scolarisation des enfants) et niveau de vie (Mesuré par le revenu national brut (RNB) par habitant, ajusté en parité de pouvoir d'achat (PPA)). Généralement, l'IDH est exprimé sous forme d'un nombre entre 0 et 1. Ainsi, le deuxième graphe **(Figure 6),** permet de montrer le classement du Maroc par rapport aux autres pays sur la même période (1990-2022).

**Figure 5 : Evolution d'Indice de développement humain du Maroc (1990-2022)**

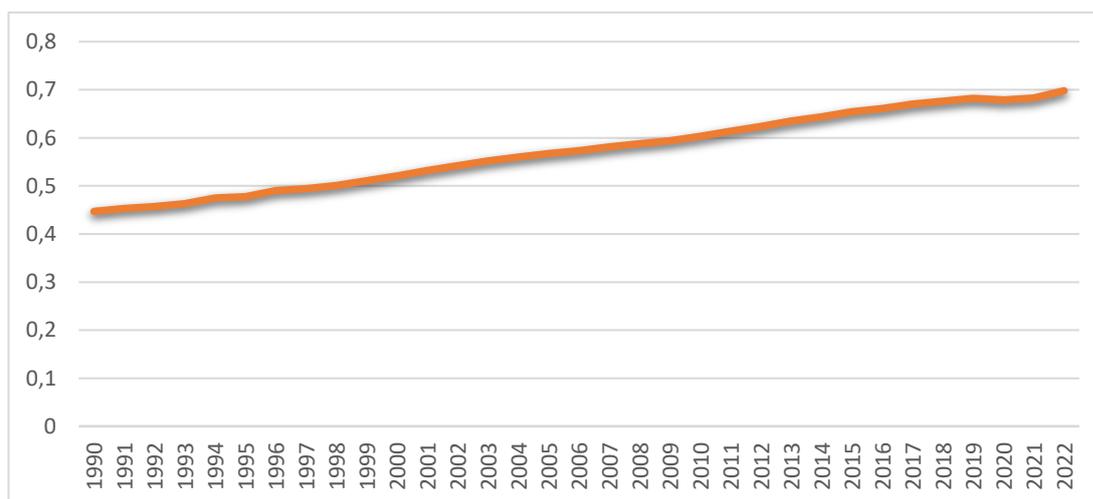

**Source : Réalisé par nos propres soins selon :  Maroc - Indice de développement humain 2021 | countryeconomy.com**

---

³ Site web : Accueil | Programme De Développement Des Nations Unies (undp.org) : hdr2023-24overviewfr.pdf (undp.org)

**Figure 6 : Classement du Maroc au niveau mondiale selon IDH (1990-2022)**

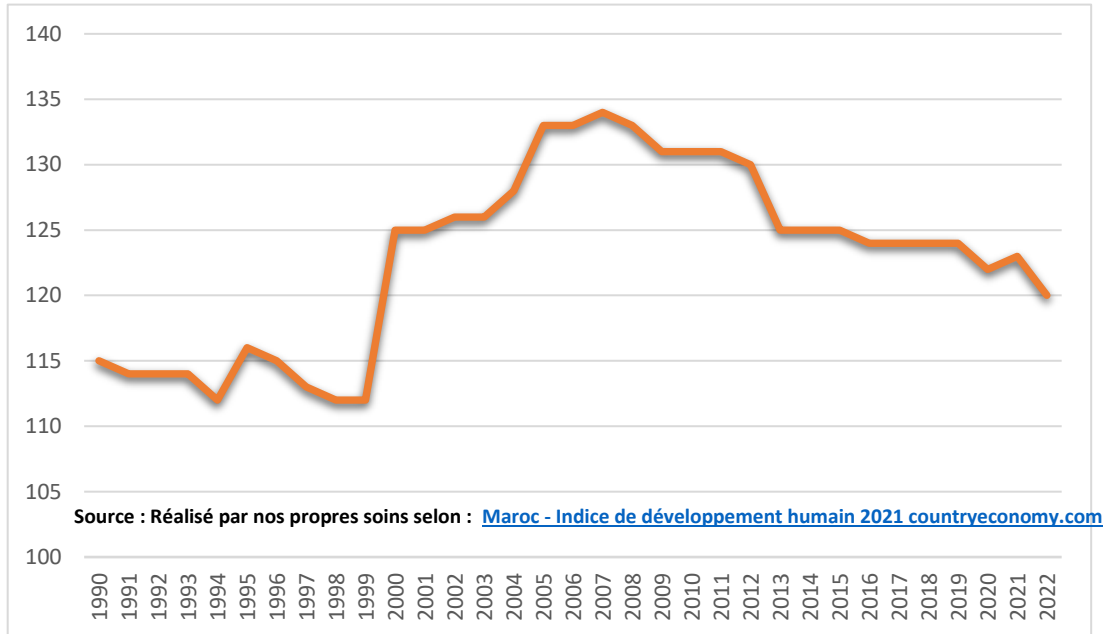

Source : Réalisé par nos propres soins selon : Maroc - Indice de développement humain 2021 countryeconomy.com

Selon l'IDH, Les pays sont répartis en quatre groupes : « Très élevé » et « élevé » représentent les pays considérés comme développés, tandis que le groupe « moyen » correspond aux pays en développement et enfin le groupe « faible » correspond aux pays en voie de développement.

Le tableau ci-après représente le classement des 4 groupes selon l'intervalle de l'IDH de chacun ainsi que son niveau :

| Niveau de développement | Intervalle IDH |
|---|---|
| **Très élevé** | 0.800-1.00 |
| **Elevé** | 0.700-0.799 |
| **Moyen** | 0.550-0.699 |
| **Faible** | 0.000-0.549 |

*Source : Selon le Rapport sur le Développement humain 2023-2024*

Selon le rapport mondial sur le développement humain 2023/2024, le Maroc a obtenu une valeur de l'IDH de 0.698 en 2022, ce qui le classe dans la catégorie de développement humain moyen.

Le Maroc a connu une amélioration dans le classement mondial de l'IDH, passant de 123$^{ème}$ à 120$^{ème}$ position mondiale. D'après l'observatoire national du développement humain (ONDH), c'est la première fois que le Maroc fait une telle progression en améliorant son classement de trois places en une seule fois. Cependant, la victoire de 3 places n'est pas suffisamment suffisante.

Le président de l'ONDH, Othman Gair[4], déclare que parmi les composantes qui ont joué un rôle dans l'amélioration de l'IDH au Maroc sont l'espérance de vie et la durée attendue de scolarisation, qui ont enregistré respectivement une augmentation de 1.4% et 4.3%. Tandis que seulement 0.1% du revenu national brut par habitant y a contribué.

## II. Engagement de la CGEM pour la promotion de la RSE :

### 1. Présentation de la CGEM :

La Confédération Général des Entreprises du Maroc (CGEM) est une association professionnelle représentant le secteur privé marocain, signataire du Pacte Mondial de l'ONU en décembre 2006 déclare **Saadia SLAOUI BENNANI** (Vice-Présidente du bureau de la CGEM et Présidente de la Commission RSE et Genre) dans le Communication sur le progrès de janvier 2020 à janvier 2022[5].

Sur le plan écologique et social, le Pacte Mondial des Nations Unies est une initiative internationale ayant pour objectif d'encourager les entreprises, associations et organisations non gouvernementales (ONG) a adopté des comportements responsables. Quatre textes fondamentaux regroupent 10 principes relatifs aux droits de l'Homme, les normes de travail, la lutte contre la corruption et l'environnement :

- Déclaration universelle des Droits de l'Homme
- Déclaration de L'OIT sur les principes du travail
- Déclaration de Rio sur l'environnement et le développement
- Convention des nation unies contre la corruption.

---

[4] Site web : Le Maroc gagne 3 rangs dans le classement mondial de l'Indice de développement humain - Médias24 (medias24.com)
[5] Site web : CGEM | La Confédération Générale des Entreprises du Maroc : CGEM - Pacte Mondial COE 2020-2021.docx

Fondée en 1947, la CGEM représente le secteur privé au Maroc. Elle compte 90.000 membres directs et affiliés dont 95% de TPME [6]. La CGEM joue un rôle de proposition auprès des pouvoirs publics car elle s'est imposée comme le représentant officiel du secteur privé auprès des partenaires sociaux, des institutions et des autorités publiques.

En sensibilisant, mobilisant et accompagnant ses membres, la CGEM joue un rôle crucial pour qu'ils prennent conscience de leur responsabilité afin qu'ils contribuent aux 17 objectifs de développement durable. En parallèle, elle est chargée de défendre les intérêts du secteur privé à l'échelle régionale, nationale et internationale tout en œuvrant à la création d'un environnement favorable et un climat propice des affaires.

La CGEM s'organise autour de 18 commissions transversales[7], dont 3 commissions veillant à la promotion de la RSE, du développement durable et de l'éthique des affaires qui sont :

| Commission RSE et diversité | Commission développement durable | Commission éthique et déontologie |
|---|---|---|
| Instituée par le Conseil d'Administration de la CGEM en 2006. Présidée par Mme. Saadia SLAOUI BENNANI et son vice-président Mme. Fatima-Zahra EL KHALIFA<br>Ses axes Stratégiques sont :<br>○ Responsabilité Sociétale des entreprises<br>○ Diversité : Promotion de la diversité sous toutes ses formes<br>○ Innovation social et entrepreneuriat social | Cette commission est présidée par Mme. Assia BENHIDA et son vice-président M. Youssef CHAQOR<br>Ses axes stratégiques :<br>○ Efficacité Hydrique<br>○ Economie circulaire<br>○ Villes et mobilité durable | Mme Chadia JAZOULI (Présidente) et Mme Hind LFAL (Vice-présidente) gouverne cette commission dont ses axes stratégiques sont :<br>○ Promouvoir la culture de la bonne gouvernance<br>○ Sensibiliser les entreprises aux politiques d'éthique : Lutte contre la corruption et lutte contre l'informel et la contrefaçon. |

*Source : Réalisé par nos soins - Inspiré du site CGEM : Commissions permanentes | CGEM*

---

[6] Site web : CGEM | La Confédération Générale des Entreprises du Maroc : Presentation-CGEM-2023-FR.pdf
[7] Site web : CGEM | La Confédération Générale des Entreprises du Maroc :Commissions permanentes | CGEM

## 2. CGEM et RSE : instauration de la Charte RSE et son Label social

La CGEM s'engage à promouvoir la RSE en encourageant l'éthique des affaires, en réponse au dynamisme des réformes économiques, sociales et environnementales qu'a connu le pays.

En 2005, elle a contribué au projet « Developpement durable grâce au pactes mondial » en collaboration avec la coopération du Bureau International de Travail (BIT) ainsi que d'autre partenaire. La CGEM a ensuite instauré une charte de RSE suivi d'un Label en 2006, afin de fournir aux entreprises des outils pour structurer leur démarche RSE.

Dans cette optique, elle a organisé le 1$^{er}$ prix visant à récompenser les entreprises socialement responsables de leurs engagements, et met en place une charte RSE ainsi que le « Label CGEM pour la RSE ». Les deux outils, élaborés sur la base des standards internationaux, permettent à la CGEM de se positionner autant qu'une organisation de référence en matière de développement durable.

La charte RSE met en évidence l'importance pour les entreprises de renforcer leurs écoute envers leurs parties prenantes et de communiquer avec la société leurs volontés et leurs ambitions de mettre l'activité économique au service du progrès humain. Les 9 domaines d'engagement identifiés dans la charte définissent chacune les objectifs des stratégies avec les orientations et lignes directrices conformément à la norme ISO 26000, puisqu'il s'agit d'une norme non certifiable mais sert de guide pour intégrer la RSE dans la stratégie et les opérations quotidiennes des organisations. Ces 9 domaines d'engagement sont à respecter par les entreprises membres de la CGEM lors de l'exécution de leurs activités et de leurs relations avec les parties prenantes.

Cependant, d'après la charte de la CGEM de RSE publié par la confédération, les 9 domaines d'engagement sont :

- Respecter les droits humains

- Améliorer en continu les conditions d'emploi et de travail et les relations professionnelles

- Préserver l'environnement

- Prévenir la corruption

- Respecter les règles de la saine concurrence

- Renfoncer la transparence du gouvernement d'entreprise

- Respecter les intérêts des clients et des consommateurs

- Promouvoir la responsabilité sociétale des fournisseurs et sous-traitant

- Développer l'engagement envers la communauté

Le Label « Label CGEM pour la RSE » est une reconnaissance des initiatives déployés par les entreprises marocaines, de leur dévouement à promouvoir les principes universels de la responsabilité sociales.

Les entreprises marocaines et membre de la CGEM reçoivent le label RSE ce qui signifie que leurs structures et leurs actions de gestion sont conformes aux objectifs de la charte RSE. Il est nécessaire de soumettre les demandes de labélisation au Président de _la commission RSE et diversité_, accompagnées d'une signature d'une personne qualifiée.

Le label aide les entreprises à définir leurs engagements, à mesurer leurs avancés et faire reconnaitre leurs performances et résultats. Cela favorise la compétitivité, facilite l'accès aux marchés, renforce la cohésion de l'équipe.

Avant toute labellisation, il est nécessaire de procéder à une évaluation, sur place et sur pièces par un tiers indépendant accrédité par la CGEM, qui est effectuée aux frais de l'entreprise. L'évaluation d'une entreprise varie en fonction de divers critères, tels que le nombre réel de ses employés et le nombre de ses sites...etc. Le comité a la possibilité d'accorder le Label si le niveau de conformité est adéquat, de le donner avec des conditions suspensives à lever ou de reporter l'examen de l'octroi du label jusqu'à correction des non-conformités observées.

Le label est valable pendant 3 ans et est délivré par le président de la CGEM au représentant qualifié de l'entreprise. Un certificat mentionnant le nom de l'entreprise est accompagné du trophée du label et remis à l'entreprise concernée lors de la cérémonie de remise du label organisé par la CGEM. Le renouvellement du label est précédé par une évaluation effectuée dans les mêmes conditions.

L'ensemble des règles d'attribution, les procédures à suivre, la charte RSE de la CGEM ainsi que les entreprises labellisées RSE sont accessibles sur le site de la Confédération : CGEM | La Confédération Générale des Entreprises du Maroc

### 3. Le Label CGEM : Une Initiative en Plein Élan.

La commission RSE et diversité a pour mission d'assoir la culture « RSE » au sein des entreprises marocaines autant que pilier inhérent à leur développement et à leur compétitivité. L'objectif est d'inciter les entreprises à tisser une relation privilégiée avec leurs parties prenantes internes et externes et de renforcer la reconnaissance internationale de la démarche RSE de la CGEM ainsi que du Label.

La CGEM s'est engagé à promouvoir la RSE auprès de ses entreprises affiliées en encourageant l'intégration des pratiques de responsabilités sociales dans leurs processus de gestion.

Le comité d'attribution du label est composé de personnalités qualifiées et reconnues dans le développement des entreprises au Maroc. Son rôle principal consiste à décider de l'attribution et du renouvellement du label RSE aux entreprises postulantes.

Notre analyse sur la tendance de labélisation des entreprises marocaines se divisera en deux phases : la première couvrira la période Pré-Covid19, tandis que la seconde se concentrera sur la période de Post-Covid19.

Selon les données communiquées et représentées sur le graphique ci-dessous, une tendance notable d'attribution et de renouvellement du label RSE par la CGEM a été identifié au cours de la période, allant de 2014 jusqu'à 2018. Cela reflète, la prise de conscience et la conviction des entreprises quant à l'importance de l'approche RSE dans le département de gestion, celle-ci étant considéré comme instrument pour favoriser l'amélioration continue, gérer les risques et de renforcer la performance globale.

**Figure 7 : Entreprises labellisées "Label RSE" par la CGEM : Période Pré-Covid 19**

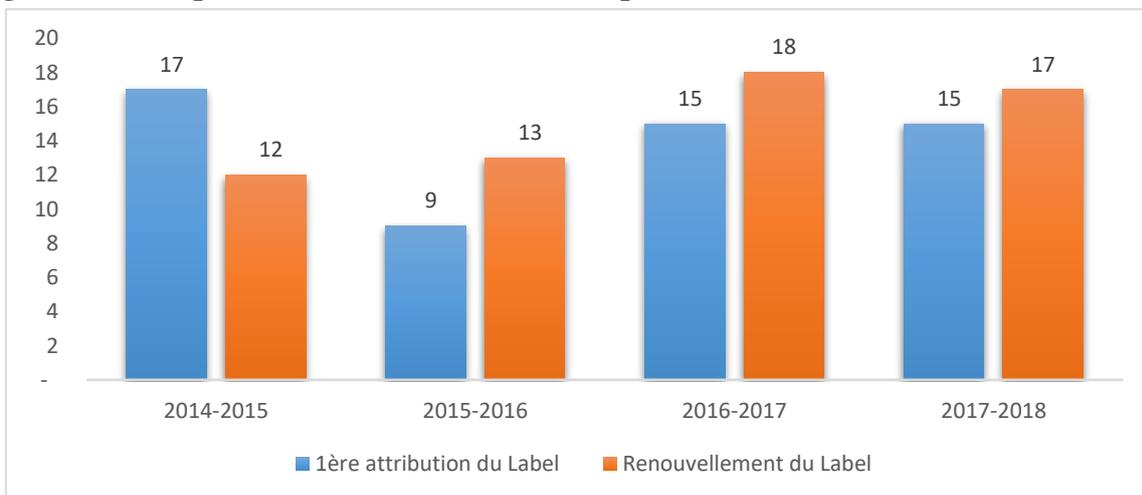

*Source : Graphique réalisé par nos propres soins - Rapport Moral de la CGEM (de 2014-2018)*

Au total, 116 entreprises sont dotées d'un label RSE de la CGEM durant cette période. L'exercice 2014-2015, s'est caractérisé par le nombre maximum de nouvelles attributions du label RSE aux entreprises marocaines. En revanche, l'exercice 2016-2017, a été marqué par une augmentation notable de renouvellement du label avec un total de 18 entreprises marocaines relabélisées. Les autres exercices manifestent également une tendance significative et témoignent l'intérêt des entreprises marocaines de poursuivre des pratiques socialement responsables afin de refléter l'image d'une bonne entreprise citoyenne exemplaire.

La crise Covid-19 a déstabilisé l'activité économique et a eu un effet sur tous les objectifs des entreprises y compris ceux relatifs au développement durable. Cette crise a posé des défis pour les grandes et petites entreprises, mettant à l'épreuve la RSE. Effectivement, la crise sanitaire a été un véritable test sur l'engagement réel et concret des entreprises sur le plan social et responsable sur le terrain.

Les entreprises ayant, effectivement, pris en compte leurs relations avec leurs parties prenantes (Salariés, fournisseurs, communautés locales…) ont pu affronter la pandémie. Tandis que ceux, ayant considéré la démarche RSE comme une simple affaire de notoriété et d'image de marque, se trouvaient davantage rencontrés à plein de difficulté.

Cependant, La RSE est perçue par les entreprises comme étant un critère incontournable d'attractivité, c'est-à-dire que l'entreprise devient plus attrayante et compétitive sur le marché en adoptant des pratiques de RSE. Son émergence est devenue une nécessité compte tenu les retombés et conséquences néfastes de la pandémie Covid-19.

Le graphe ci-dessous, met en évidence une tendance significative des entreprises marocaines disposant du label pour la 1$^{ère}$ fois ou le renouvelant au cours de la période Post-Covid-19.

**Figure 8 : Entreprises labellisées "Label RSE" par la CGEM : Période Post-Covid 19**

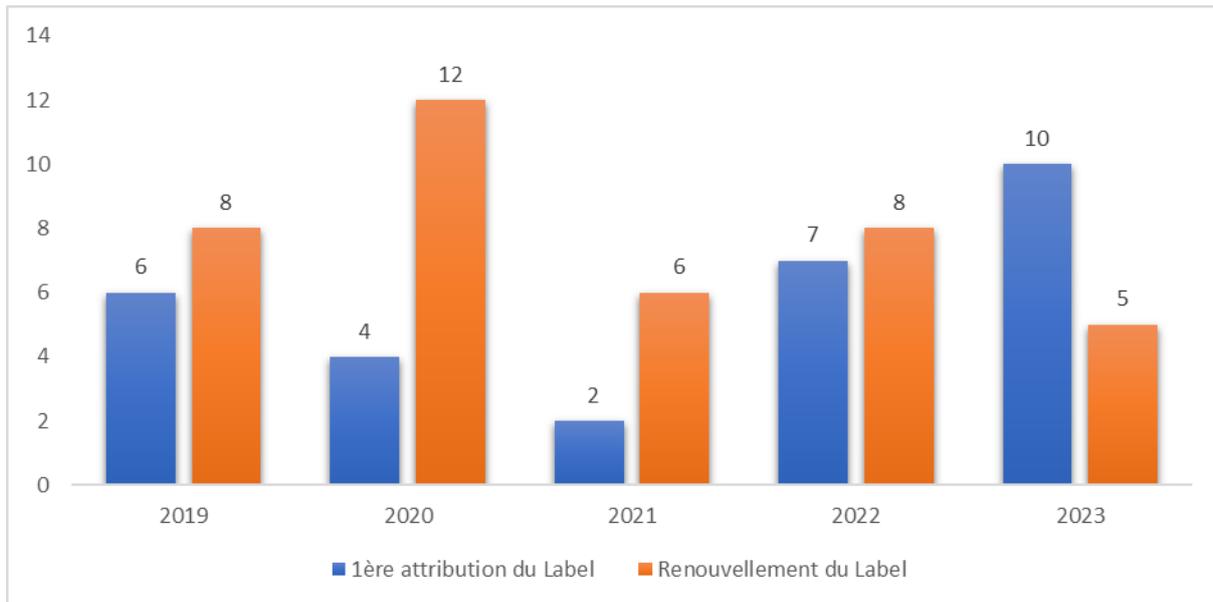

*Source : Graphique réalisé par nos propres soins - Rapport Moral de la CGEM (de 2019-2023)*

Après la pandémie de Covid 19, de nombreuses entreprises ont été incitées à s'orienter vers l'adoption des pratiques socialement responsables. En 2020, on observe une augmentation significative du nombre d'entreprises renouvelant le label RSE, avec un total de 12 entreprises marocaines, avec 4 entreprises ayant obtenu ce label pour la première fois. En revanche, en 2023, le label a été attribué à 10 nouvelles entreprises, avec 5 autres entreprises renouvelant la certification.

Cependant, la période Post-Covid 19 est marquée par une tendance aussi bien significative, suggérant que la labélisation RSE des entreprises marocaines est dans une allure croissante d'après les données communiquées par la CGEM et représenté sur les graphes.

L'engagement accru des entreprises dans l'obtention du label, ne reflète que leurs volontés d'implication profonde dans le terrain social afin de projeter une image d'entreprise citoyenne, de renforcer des liens avec l'ensemble des parties prenantes et de bénéficier d'un avantage concurrentiel. La pandémie Covid19 a agi comme un catalyseur augmentant la prise de conscience accrue de l'importance de la RSE.

Actuellement, nous comptons 124 entreprises marocaines[8] de divers secteurs et tailles, bénéficiant de cette distinction. L'objectif est d'inciter les entreprises à intégrer dans leurs stratégies les pratiques de la RSE pour une meilleure résilience future.

❖ **Ventilation des entreprises labelisées RSE par secteurs d'activité :**

Selon, le CoP (Communication sur le progrès) de la période de janvier 2020 à janvier 2022, indique que durant les deux années 2020 et 2021, les entreprises labellisées, au nombre de 108, sont ventilées, selon le secteur d'activité, comme suit :

**Figure 9 : Entreprises labellisées RSE par secteur d'activité (2020-2021)**

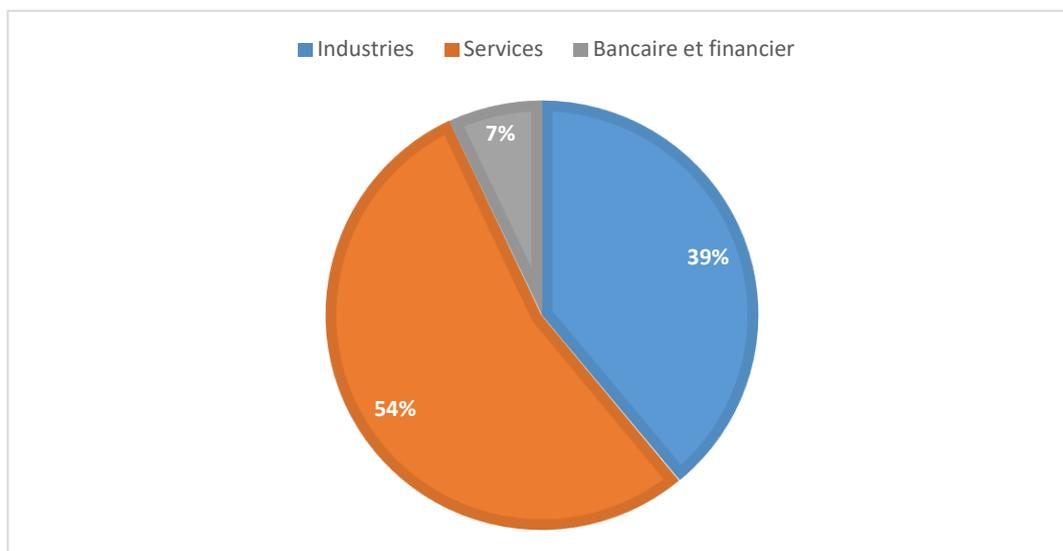

*Source : Graphique réalisé par nos propres soins- PACTE MONDIAL Des Nations Unies CoP (Janvier 2020-Janvier 2022)*

D'après le graphique, nous constatons que les entreprises labellisées RSE opérant dans le secteur de service représentent une part significative, atteignant 54% du total pour la période 2020-2021. Par ailleurs, 39% des entreprises labellisées RSE opérant dans le secteur d'industrie, tandis que le secteur bancaire et financier ne représente que 7% du total.

Ainsi, selon le CoP de la période de janvier 2022 à janvier 2024, nous informe sur la répartition des 124 entreprises distinguées par la labélisation en fonction du secteur pour la période 2022-2023, comme le montre le graphique suivant :

---

[8] Site web : Homepage | UN Global Compact : Confederation Generale des Entreprises du Maroc - CGEM – Communication on Engagement | UN Global Compact

**Figure 10 : Entreprises labellisées RSE par secteur d'activité 2022-2023**

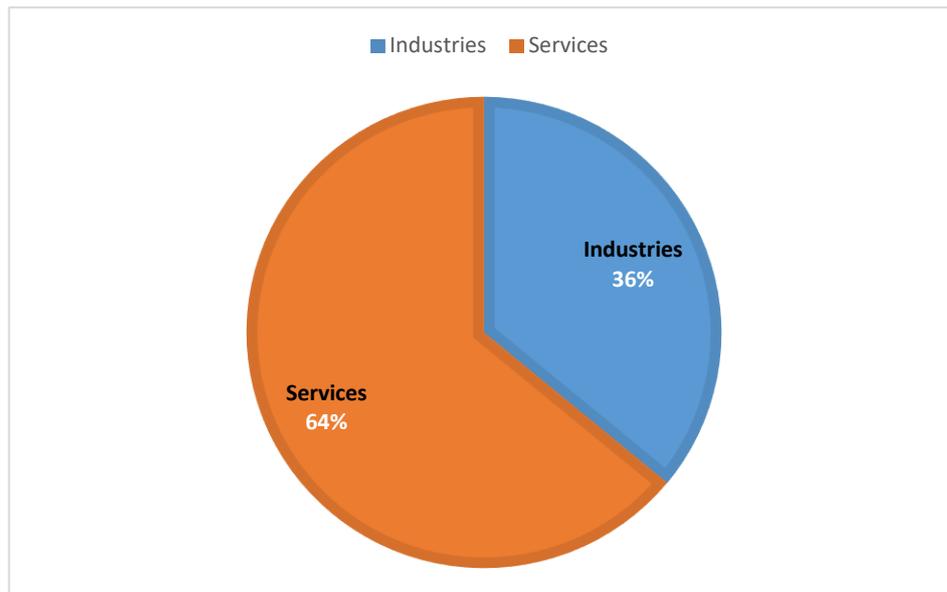

*Source : Graphique réalisé par nos propres soins - PACTE MONDIAL Des Nations Unies CoP (Janvier 2022-Janvier 2024)*

L'analyse du graphe révèle que parmi les secteurs d'activité, celui des services détient toujours la part majoritaire et représente 64% du total. Cela indique, que les entreprises qui s'orientent et choisissent de se tourner vers la labélisation RSE par la CGEM opèrent principalement dans ce secteur. Ainsi, le secteur de l'industrie représente les 36% restants du total. La dominance du secteur de service peut être expliqué par plusieurs facteurs.

Les entreprises opérant dans le secteur de service s'orientent plus vers la labellisation RSE du fait que leurs activités ont moins d'impact direct sur l'environnement en matière de consommation des ressources naturelles. Tandis que celles industrielles sont souvent confrontées à des défis environnementaux significatifs tels que pollution, gestion de déchets…etc. De surcroit, la flexibilité et la capacité d'innovation des entreprises de services leur permettent d'adopter plus facilement les pratiques RSE. Ceci se manifeste par la mise en place des politiques à savoir le télétravail, la formation continue ou les programmes de bien-être pour les collaborateurs, sans être entravées par des contraintes de processus de production industrielle, contrairement aux entreprises industrielles.

Selon une enquête nationale auprès des entreprises réalisée par le HCP en 2019[9], la répartition des entreprises par tailles et secteurs d'activité montre que plus de deux tiers des entreprises opèrent dans le secteur tertiaire. En 2019, 43.5% des TPE, 41.5% des PME, et 23.8% des GE

---

[9] Site web  Site institutionnel du Haut-Commissariat au Plan du Royaume du Maroc (hcp.ma) : E (hcp.ma)

opèrent dans le secteur de service. En revanche, seulement, 7.7% de TPE, 10.9% de PME et 25.8% des GE opèrent dans le secteur de l'industrie. La création des entreprises se concentre donc principalement dans le secteur tertiaire, tandis que le secteur de l'industrie connait de moindre croissance.

Selon l'OMPIC en 2022, les intentions de création d'entreprises qui s'élèvent à 118 622 noms commerciaux, sont réparties selon leurs secteurs d'activité comme suit :

| **Services** | **27.3%** |
|---|---|
| Commerce | 19.3% |
| BTP et activités immobilières | 19% |
| **Industries** | **10%** |
| Hôtels et restaurants | 9% |
| Transport | 6% |
| Secteur TIC | 4% |
| Agriculture et pêche | 3.2% |
| Activité financières | 2.2% |

*Source : OMPIC- LES CHIFFRES DE LA CRÉATION D'ENTREPRISES EN 2022[10]*

La répartition représentée dans le tableau ci-dessus met en lumière la prépondérance du secteur de service avec un pourcentage de 27.3%. En revanche le secteur de l'industrie ne représente que 10%. Ceci reflète l'orientation des entreprises vers des activités de prestation de services qu'elles soient professionnelles, personnelles, administratives…etc.

Les entreprises tournent vers le label pour témoigner de leur intérêt pour les pratiques socialement responsables et de leur volonté d'adopter des nouvelles démarches qui s'alignent avec les exigences du terrain que ce soit sur le plan social et/ou environnemental.

❖ **Ventilation des entreprises labelisées RSE par taille d'entreprise :**

---

[10]Site web : Office Marocain de la Propriété Industrielle et Commerciale | (ompic.ma) :  Les Chiffres de la Création d'Entreprises en 2022 | Office Marocain de la Propriété Industrielle et Commerciale (ompic.ma)

D'après les communications sur le progrès publié dans le Pacte mondiale des nations unies, les entreprises labellisées RSE par la CGEM peuvent être réparties en fonction de leurs tailles, comme montre le graphique ci-dessous :

**Figure 11 : Entreprises labelisées RSE par taille d'entreprise :**

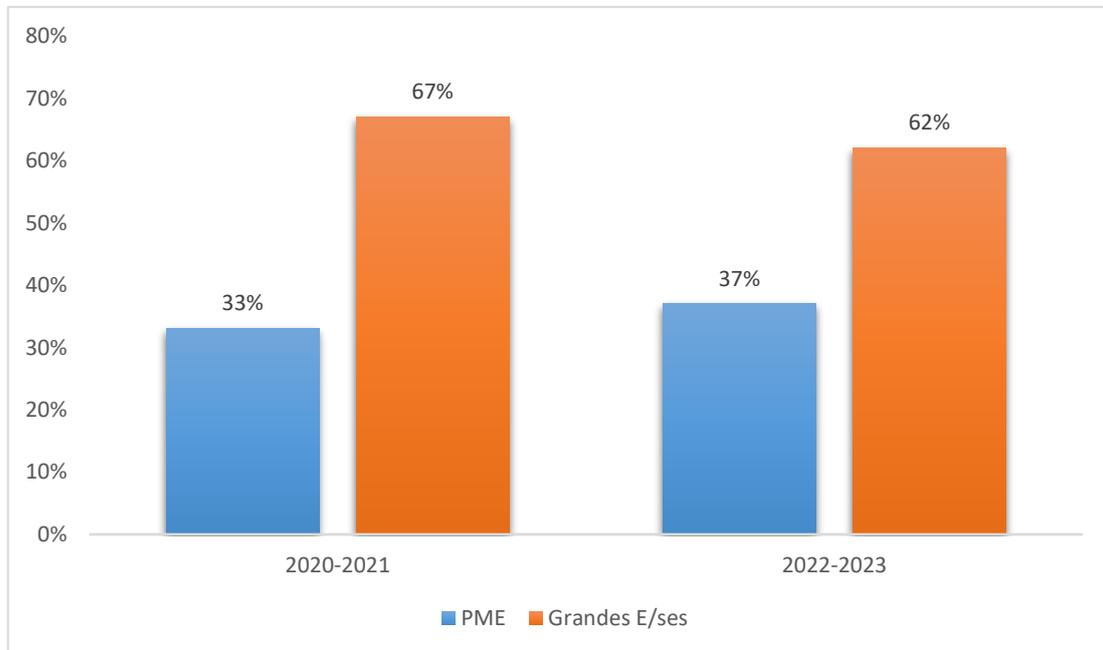

*Source : Graphique réalisé par nos propres soins- PACTE MONDIAL Des Nations Unies : CoP (Janvier 2020-Janvier 2022) et CoP (Janvier 2022-Janvier 2024)*

Selon le rapport annuel 2021-2022 de l'Observatoire Marocain de la Très Petite et Moyenne Entreprises (OMTPME), les TPME représentent 99.6% du tissu économique marocain.

En 2020-2021, 33% des entreprises labellisées RSE étaient des petites et moyennes entreprises (PME). Leur proportion a augmenté au cours de l'exercice 2022-2023, pour représenter 37% du total. Les grandes entreprises, quant à elles, représentaient la majorité des labélisations RSE avec 67% et 62% pour respectivement l'exercice 2020-2021 et 2022-2023, marquant ainsi une baisse de 5%.

Il s'agit de noter qu'en terme d'évolution d'entreprise entre 2020 et 2021, 17 233 [11]entreprises ont changé de catégorie, précise Idrissi Amal, directrice exécutive de l'OMTPME. Les statistiques montrent que 7 360 microentreprises sont devenues des TPE, 2 563 sont passées de TPE en PME, et 229 PME sont passées à la catégorie de grandes entreprises (GE). Tandis que,

---

[11] Site web : OMTPME: les TPME représentent 99,6% du tissu économique marocain (fnh.ma)

51 grandes entreprises sont passées à la catégorie de PME, 1 505 PME des TPE ou des microentreprises et 3 266 TPE des microentreprises.[12]

L'augmentation de l'adoption des pratiques RSE par les PME et la baisse de celles des grandes entreprises en 2022-2023, peut-être expliqué par plusieurs facteurs. D'une part, le changement récent de la catégorie des entreprises, passant de grandes entreprises déjà labellisées ou potentielles candidates à la labellisation en raison de l'adéquation de leurs conditions aux critères du label, à des petites entreprises. D'autre part, il est important de souligner l'impact économique et structurel lié à la pandémie de Covid19. Celle-ci, a entrainé une chute significative de la demande pour une panoplie de produits et services, forçant les grandes entreprises à diminuer leurs activités. Ainsi, les pertes financières ont poussé certaines grandes entreprises à restructurer ou à se scinder en unités petites et agiles pour survivre. Par ailleurs, les modifications et changements des stratégies ou de propriété au sein de ces entreprises par la réorientation de leurs efforts vers d'autres initiatives de durabilité, contribuent également à expliquer l'évolution des pourcentages d'entreprise labellisées RSE.

❖ **Cotation des entreprises labelisées RSE :**

Actuellement, les 23 entreprises labellisées RSE représentent environ 30% des 76 entreprises cotées à la Bourse de Casablanca[13]. Cela reflète une adoption notable des pratiques de RSE parmi les sociétés cotées démontrant et indiquant un engagement significatif envers la durabilité et les pratiques éthiques.

Il est important de noter que lors du dernier CoP 2020-2022, il y'avait 24 entreprises labellisées RSE cotées. Cependant, pour le CoP 2022-2024, ce nombre est passé à 23, ce qui indique qu'une entreprise labellisée a quitté la bourse. Malgré cette légère baisse numérique, la proportion relative des entreprises labellisées RSE parmi toutes les entreprises cotées en bourse représente 30% et demeure toutefois significative.

Cette évolution met en lumière l'importance croissante accordée à la RSE par les entreprises cotées en Bourse de Casablanca, ainsi que la reconnaissance de son influence positive sur la réputation, durabilité des opérations commerciales et les relations avec les parties prenantes.

---

[12] Site web : Page d'accueil - L'Observatoire Marocain de la TPME - OMTPME : Rapport-consolide-25.10.-VF.pdf (omtpme.ma)
[13] Site web : Bourse de Casablanca - La Bourse Pour Tous (casablancabourse.com)

- ❖ **Ventilation des entreprises labelisées par leurs structures de propriété et leurs origines :**

Les entreprises « Maroco-marocaines » sont fondées, enregistrées et fonctionnent principalement au Maroc, ce qui les ancre profondément dans le tissu économique local et répondant aux besoins du marché marocain. Elles sont généralement indépendantes de toute influence ou contrôle étranger, contrairement aux filiales de multinationales qui sont des extensions de sociétés basées à l'étranger.

Selon les CoPs de la période de 2020-2022 et 2022-2024, 32% des entreprises labellisées RSE sont des filiales internationales, tandis que 68% des entreprises sont des maroco-marocaines. Cette répartition démontre l'adoption volontaire des standards éthiques contribuant ainsi à un développement durable. Cela témoigne également un fort engagement local envers les pratiques de durabilités améliorant la performance sociale et environnementale au niveau national.

**Conclusion :**

Les entreprises marocaines deviennent de plus en plus conscientes de l'importance d'intégrer une démarche RSE dans leurs stratégies internes et externes en tenant compte l'ensemble de ses parties prenantes. Elles cherchent à l'appliquer non seulement sous un cadre mercatique pour des raisons de réputation et image de marque, mais plutôt une intégration effective en respectant et adoptant les pratiques socialement responsables.

L'adoption de la démarche RSE reste volontaire pour les entreprises. Aucun texte ou loi marocaine, ne les imposent à la mettre en œuvre. Cependant, les entreprises comprennent que face aux récentes réformes économiques, sociales et environnementale qu'a connue le Maroc, l'obligation de s'aligner avec les exigences du pays est devenue impérative. Un développement sain et durable de leurs activités ne peut être envisagé que dans un environnement prospère.

La Confédération Générale des Entreprises au Maroc a lancé en 2006 deux initiatives importantes : La charte RSE et le Label CGEM pour la RSE. La charte RSE joue le rôle d'un guide pour les entreprises dans le but de les encourager à adopter des pratiques socialement responsables et est organisée autour de 9 domaines d'engagement. Parallèlement, le label CGEM est une distinction par laquelle la CGEM reconnait l'implication des entreprises dans la démarche RSE. Il sert également d'outil pour évaluer et mesurer la performance sociale des entreprises ainsi que de suivre leurs avancements dans l'intégration de la RSE dans leurs stratégies.

D'après le rapport de Communication sur le Progrès (CoP) de 2022-2024 du Global Compact, nous comptons 124 entreprises marocaines labellisées RSE. Pour un label existant depuis 2006, ce nombre témoigne d'une certaine réticence des entreprises marocaines à mettre en œuvre pleinement la démarche RSE dans leurs stratégies, tant internes qu'externes. Cette hésitation peut être attribuée à la faible pression de la société civile pour inciter les entrepreneurs à changer de mentalité et adopter la RSE non seulement pour des raisons d'images ou réputation courtermiste mais également comme un moteur potentiel de croissance et un levier de développement. De surcroit, selon Filali Maknassi (2009) la RSE reste souffrante de quelques entraves qui sont la qualification du personnel, les ressources financières et le manque d'information.

Le Maroc est au cœur d'un processus de transition en incluant et intégrant de nouvelles pratiques au sein du tissu économique marocain, mettant l'accent sur des pratiques sociales, sociétal et environnementale. Cette transition reflète un engagement philanthropique en faveur du développement durable et la promotion de la RSE.

**Référence Bibliographique :**